\documentclass[aps,pra,epsfigure,twocolumn,showpacs]{revtex4}
\usepackage{graphicx}
\usepackage{amsmath}
\usepackage{latexsym}
\usepackage{amsfonts}
\usepackage{amssymb}
\usepackage{array}
\usepackage{epsfig}

\newcommand{\ket}[1]{\left\vert#1\right\rangle}

\begin{document}

\title{   
Testing Bell inequalities 
with photon-subtracted Gaussian states
}
\date{\today}

\author{Hyunseok Jeong$^{1,2}$}

\affiliation{
$^1$School of Physics and Astronomy, Seoul National University,
Seoul 151-747, Korea\\
$^2$Center for Quantum Computer Technology, Department of Physics,
University of Queensland, Brisbane, Qld 4072, Australia
}
\date{\today}

\begin{abstract}
Recently, photon subtracted Gaussian states (PSGSs) were generated by several
experimental groups. Those states were called ``Schr\"odinger kittens"
due to their similarities to superpositions of coherent states (SCSs) with
small amplitudes. We compare the ideal SCSs and the PSGSs for 
experimental tests of certain types of Bell inequalities.
In particular, we analyze the effects of the key 
experimental components used to generate PSGSs:
mixedness of the Gaussian states, limited transmittivity of the
beam splitter and the avalanche photodetector which cannot
resolve photon numbers. As a result of this analysis, the degrees
of mixedness and the beam splitter transmittivity that can be allowed
for successful tests of Bell inequalities are revealed.    
\end{abstract}
\pacs{03.67.Mn, 42.50.Dv, 03.65.Ud, 42.50.-p}

\maketitle

\section{introduction}

Quantum optics enables one to experimentally
explore quantum physics and
quantum information processing
in the context of continuous variables.
Gaussian continuous-variable states
with nonclassical properties,
such as single-mode and two-mode squeezed states
have been generated and used 
for various applications
in numerous quantum optics experiments.
For example,
continuous-variable quantum teleportation 
has been performed using the two-mode
squeezed states of light
\cite{Furusawa98,Bowen03}.
However, it is experimentally more difficult to
generate and control
non-Gaussian continuous variable states.

Superpositions of two coherent states
in free-traveling optical fields (SCSs),
a well known class of non-Gaussian continuous variable states,
have attracted remarkable attention.
The SCSs, when the component coherent states are well separate
in the phase space, are often called 
``Schr\"odinger cat states'' 
as they show typical properties
of macroscopic quantum superpositions \cite{Schr,WScat}.
The SCSs show nonclassical properties such as
interference patterns in the phase space
and negative values in the Wigner functions
\cite{WScat}.
Once single-mode SCSs are generated, it is relatively easy to
generate two-mode cat states
using a 50:50 beam splitter. These two-mode
SCSs (also called entangled coherent states \cite{Sanders})
have been found
\cite{Derek,jeongsonkim,Magda} to 
violate Bell inequalities \cite{Bell,CHSH,CH}
using various types of measurements.
It has also been shown that the free-traveling SCSs can be
useful for various applications in quantum information processing (QIP)
\cite{Enk01,JKL01,Jeong02,Ralph03,WeakForce}.
All the QIP applications 
and Bell inequality tests using SCSs
require free-traveling SCSs with reasonably large amplitudes.
For example, Lund {\it et al.} showed that SCSs with amplitudes $\alpha> 1.2$ 
allow for fault tolerant quantum computing
with reasonable photon loss \cite{Lund07}.

On the other hand, until recently, 
it has been known to be extremely hard to experimentally
generate free-traveling SCSs. There have been
schemes to generate such SCSs using  
strong nonlinear interactions \cite{Yurke}
or photon number resolving detectors
 \cite{Dakna,Dakna2},
neither of which is feasible using current technology.
Recently, 
the generation of free-traveling SCSs have been studied
by several authors and
more feasible schemes have been suggested
\cite{Lund04,jc1,jc2,jc3,jc4,KM}. 
Among those, a simple and useful observation was made that  
SCSs with small amplitudes, such as $\alpha<1.2$,
are very well approximated by squeezed single photons
\cite{Lund04}.
It was also pointed out that squeezed single photons are identical
to the PSGSs obtained by subtracting one photon from pure
squeezed vacuums \cite{JLR05}.
Meanwhile, photon subtracted Gaussian states (PSGSs) were
generated by several experimental groups
\cite{kt1,kt2,kt3,kt4}.
Those states were called ``Schr\"odinger kittens"
as they are close to the SCSs 
with small amplitudes ($\alpha< 1.2$).
Some effects of experimental imperfections including
inconclusive photon subtraction and impurity of Gaussian sources 
were theoretically analyzed \cite{Kim05,OP}.
It was pointed out that subtracting a single photon
from a Gaussian state is experimentally more efficient than
directly squeezing the single-photon state \cite{SS06}.

Very recently, a remarkable breakthrough was made:
free-traveling SCSs, now called ``Schr\"odinger cats",
were generated and detected \cite{Cat07}, where the size of the states
($\alpha=1.6$) were reasonably large for fundamental tests
of quantum theory and quantum information processing.
However, the fidelity of the generated states is yet to be improved
for practical quantum information processing.
In the meantime, the PSGSs will be useful for small-scale tests of
Bell inequalities and quantum information processing.
Experimental efforts and progress are being made
to generate SCSs of even larger amplitudes and higher fidelity
\cite{Sasaki}.

In this paper, we study tests of certain types of 
Bell inequalities \cite{BW} with
the PSGSs and discuss the possibility of experimental
realization using current technology.
Gaussian squeezed states, a beam splitter 
and an avalanche photodetector are
experimental components typically used
to generate PSGSs \cite{kt1,kt2,kt3,kt4}.
We analyze the detrimental effects of these  
experimental components, i.e.,
mixedness of the Gaussian states,
limited transmittivity of the beam splitter 
and the avalanche photodetector which cannot
resolve photon numbers.
As a result of this analysis, we reveal 
the degrees of mixedness and the beam
splitter transmittivity that can be allowed
for successful tests of Bell inequalities.
In particular, we show that quantum nonlocality may be
verified using current technology with
homodyne detection based on the
Clauser and Horn (CH)'s version of Bell's inequality \cite{CH}.

There have been publications on the study of Bell's inequality 
utilizing photon subtraction methods including Refs.~\cite{OP05,In05,FP,Nha,G04}. 
In particular, a state generated by subtracting 
one photon from each mode (total two photons)
of a two-mode squeezed state
was considered for Bell inequality tests~\cite{OP05,In05,FP,Nha,G04}.
We note that while such two-photon subtraction
is useful to enhance quantum
nonlocality~\cite{OP05,In05,FP,Nha,G04},
the method considered in our paper using
the PSGS can be %experimentally
more efficient ({\it i.e.} the success probability is larger)
because it require only a single-photon subtraction.

This paper is organized as follows. 
In Sec.~II, we briefly review 
the comparison between the pure PSGS and the SCS
by means of optimized fidelity \cite{Lund04}.
In Sec.~III, we investigate Bell inequality tests
suggested by Banaszek and W{\' o}dkiewicz (BW) \cite{BW}
for ideal PSGSs and compare the results with those of 
ideal SCSs.
In Sec.~IV, we consider experimental detrimental effects
based on the approaches in Refs.~\cite{Kim05} and \cite{OP}
for our Bell inequality tests and analyze
the results.
We then conclude the paper in Sec.~V
 with final remarks on
prospects of experimental tests of Bell
inequalities using PSGSs.

\section{Superposition of coherent states and
photon subtracted squeezed vacuum}

A SCS can be represented as
\begin{equation}
\label{CSSdefine}
|{\rm SCS}_\varphi(\alpha)\rangle=
N_\varphi(\alpha)(|\alpha\rangle+e^{i\varphi}|-\alpha\rangle),
\end{equation}
where $N_\varphi(\alpha)$ is a normalization factor,
$|\pm\alpha\rangle$ is a coherent state of amplitude $\pm\alpha$,
and $\varphi$ is a real local phase factor. 
The amplitude $\alpha$ is assumed to be real for simplicity
without loss of generality.  
The size of the SCS can be defined by the magnitude of the amplitude $\alpha$.  
The SCSs such as 
$|{\rm SCS}_\pm(\alpha)\rangle
=N_{\pm}(\alpha)(|\alpha\rangle\pm|-\alpha\rangle)$
are called even and odd SCSs respectively because the even (odd) SCS
always contains an even (odd) number of photons.
By splitting a SCS at a 50:50 beam splitter, two-mode entangled coherent state 
\begin{equation}
\label{ECS}
|{\rm ECS_\varphi(\alpha)}\rangle=
N_\varphi(\alpha)(|\beta\rangle
|\beta\rangle+e^{i\varphi}|-\beta\rangle|-\beta\rangle),
\end{equation}
where $\beta=\alpha/\sqrt{2}$, can be simply generated.

Jeong {\it et al.} pointed out that a squeezed single photon is 
identical to a {\it pure}
 PSGS ({\it i.e. an exact single photon subtraction from
 a pure Gaussian state}) \cite{JLR05}.  
This can be shown by applying the annihilation operator ${\hat a}$ to a squeezed
vacuum state: 
\begin{equation}
\hat a S(r)|0\rangle=-\sinh r S(r)|1\rangle
\label{sspm}
\end{equation}
where  $S(r)= e^{(r/2)(\hat{a}^2 - \hat{a}^{\dagger 2})}$ is
the single-mode squeezing operator and the squeezing parameter $r$
is supposed to be real throughout the paper. 
We note that a squeezed single photon
can also be obtained by adding a photon to
a squeezed vacuum as
$\hat a^\dagger S(r)|0\rangle=\cosh r S(r)|1\rangle$.

\begin{figure}
\centerline{\scalebox{0.75}{\includegraphics{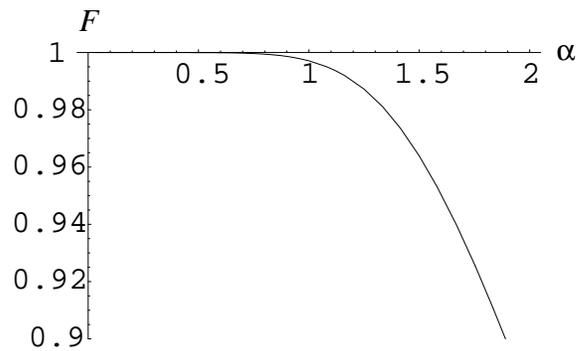}}}
\vspace{0.3cm}
\caption{
The maximized fidelity $F$ between an ideal odd cat
of amplitude $\alpha$ and a pure photon subtracted Gaussian state.
}
\label{fig:fid}
\end{figure}

Lund {\it et al.} showed that a small odd SCS with $\alpha \leq 1.2$ is
well approximated by a squeezed single photon (i.e. a pure PSGS)
as follows \cite{Lund04}.
When the squeezing operator is applied to a single photon the resultant state
can be expanded in terms of photon number states as
\begin{equation}
S(r) \ket{1} =
\sum_{n=0}^{\infty} \frac{(-\tanh r)^n}{(\cosh r)^\frac{3}{2}} 
\frac{\sqrt{(2n+1)!}}{2^n n!} \ket{2n + 1}.
\label{e-fs}
\end{equation}
The fidelity of this state to an odd SCS is
\begin{equation}
F(r,\alpha) = |\langle {\rm SCS}_-(\alpha)|S(r)|1\rangle|^2
=\frac{2\alpha^2\exp[-\alpha^2(\tanh r+1)]}{(\cosh r)^3(1-\exp[-2\alpha^2])}.
\nonumber
\end{equation}
Fig.~\ref{fig:fid} shows the maximized fidelity on the y-axis plotted
against a range of possible values for $\alpha$ for the desired odd SCS.
It shows that the fidelity approaches unity
for $\alpha$ very close to zero, while 
it decreases as $\alpha$ gets larger.
The fidelity is maximized when $r$ satisfies
\begin{equation}
\cosh r = \sqrt{\frac{1}{2} + \frac{1}{6} \sqrt{9 + 4\alpha^4}}.
\label{max_cond}
\end{equation}
The fidelity is $F>0.99$ when $\alpha<1.2$.
Some example values are:
  $F=0.9998$ for $\alpha=1/\sqrt{2}$ 
and  $F=0.997$ for $\alpha=1$, where the maximizing squeezing parameters are 
 $r=-0.164$ and $r=-0.313$ respectively.

\section{violations of Bell's inequality with pure states}

\subsection{Bell-CHCH inequality formalism with the Wigner functions}

Banaszek and W{\' o}dkiewicz (BW)  studied 
Clauser, Horne, Shimony and Holt (CHSH)'s 
version of Bell's inequality 
 based upon photon number 
 parity measurements
\cite{BW}. 
The measurement operator is defined as
\begin{equation}
\begin{aligned}
&\Pi(z)=\Pi^+(z)-\Pi^-(z) \\
&~~~~~~=D(z)\sum_{n=0}^\infty\Big(|2n\rangle\langle 2n|
  -|2n+1\rangle\langle 2n+1|\Big)D^\dagger(z)
\label{eq:bw}
\end{aligned}
\end{equation}
where $D(z)$ is the displacement operator $D(z)=
\exp[z \hat{a}^\dagger-z^* \hat{a}]$ for bosonic operators $\hat{a}$ and 
$\hat{a}^\dag$. 
Such a measurement should be able to discriminate between even numbers and
odd numbers of photons.
It is known that the displacement operation can be effectively performed using a beam
splitter with the transmittivity close to one and a strong
coherent state being injected into the other input port. 
It was pointed out that in order to maximize the violation of
the Bell-CHSH inequality for
 two-mode squeezed states and entangled coherent states, 
the BW formalism needs to be generalized to write the Bell operator as
\cite{jeongsonkim}
\begin{eqnarray}
&&{\cal B}_{BW}=\Pi_1(z_1)\Pi_2(z_2)+\Pi_1(z_1^\prime)\Pi_2(z_2)
\nonumber\\
&&~~~~~~~~~~~~~~+\Pi_1(z_1)\Pi_2(z_2^\prime)-
\Pi_1(z_1^\prime)\Pi_2(z_2^\prime)
\end{eqnarray}
while BW assumed two of the four parameters equal to zero as $z_1=z_2=0$.  
The Bell-CHSH inequality can then be represented by the Wigner function as
\begin{equation}
\begin{aligned}
&|B_{CHSH}|=|\langle{\cal
  B}_{BW}\rangle|\\
&=\frac{\pi^2}{4}|W(z_1,z_2)+W(z_1,z_2^\prime)
%\nonumber\\
%&~&~~~~
+W(z_1^\prime,z_2)-W(z_1^\prime,z_2^\prime)|\leq2,
\label{eq:BF}
\end{aligned}
\end{equation}
where $W(z_1,z_2)$ represents the Wigner function of a given
state. 
In this paper, we refer to $B_{CHSH}$ as the Bell-CHSH function.
 Using $\Pi_1(z_1)\Pi_1(z_1)=\Pi_2(z_2)\Pi_2(z_2)=\openone$,
it is straightforward to check
 the Cirel'son bound $|\langle{\cal B}_{BW}\rangle|\leq
2\sqrt{2}$ \cite{C80}
in the generalized BW formalism \cite{jeongsonkim}.

The Wigner function of a PSGS can be obtained
from its characteristic function 
\begin{equation}
\begin{aligned}
&\chi_s(\eta)={\rm Tr}\Big[S(r)|1\rangle\langle 1|
S^\dagger(r)e^{\eta {\hat a}^\dagger
-\eta^*{\hat a}}\Big]\\
&~~~~
=\exp\big[-\frac{1}{2}(e^{2r}\eta_r^2+e^{-2r}\eta_i^2)
\big](1-e^{2r}\eta_r^2+e^{-2r}\eta_i^2).
\end{aligned}
\end{equation}
The Wigner function is then
\begin{equation}
\begin{aligned}
&W_s(z)=\frac{1}{\pi^2}\int 
e^{\eta^*z-\eta z^*}\chi_s(\eta)d^2\eta\\
&~~~
=\frac{2}{\pi}\exp[-2(e^{2r }z_r^2+e^{-2r }z_i^2)]
(4e^{2r}z_r^2+4e^{-2r}z_i^2-1).
\label{koala}
\end{aligned}
\end{equation}
In order to perform a Bell inequality test,
the single-mode PSGS should be divide by a beam splitter
to generate a two-mode state shared by distant parties.
The beam-splitter operator ${\hat{O}_{BS}}$ acting on
 modes $\hat a$ and $\hat b$ is represented as 
\begin{equation}
\hat{O}_{BS}(\theta)=\exp \{\frac{\theta}{2}
(\hat{a}^{\dagger }\hat{b}
-\hat{b}^{\dagger }\hat{a})\},
\end{equation} 
where the reflectivity and transmmittivity are defined as 
$R=\sin^2(\theta /2)$ and $T=1-R$, respectively.
When the PSGS passes through a 50:50 beam splitter,
the resulting state is
\begin{equation}
W_{tot}(z_1,z_2)=W_s(\frac{z_1+z_2}
{\sqrt{2}})W_v(\frac{-z_1+z_2}{\sqrt{2}})
\end{equation}
where $W_v(z)$ is the Wigner function of the vacuum
\begin{equation}
W_v(z)=\frac{2}{\pi}\exp[-2|z|^2].
\end{equation}
The two-mode state $W_{tot}(z_1,z_2)$ can be used 
to calculate the Bell-CHSH function in Eq.~(\ref{eq:BF}).
The Wigner function of the SCS is obtained by the same method as 
\begin{equation}
W_c^\pm(z)=\frac{e^{-2|z|^2}}{\pi(1\pm e^{-2\alpha^2})}
\Big\{ e^{-2\alpha^2}(e^{-4\alpha z_r}+e^{4\alpha z_r})
\pm 2\cos 4\alpha z_i \Big\},
\end{equation}
where $W_c^+(z)$ ($W_c^-(z)$) is the Wigner function of the even (odd) SCS.
The Bell-CHSH function can be obtained in the same manner. 

\begin{figure}
\centerline{(a)~~\hspace{.32cm}
\scalebox{0.75}{\includegraphics{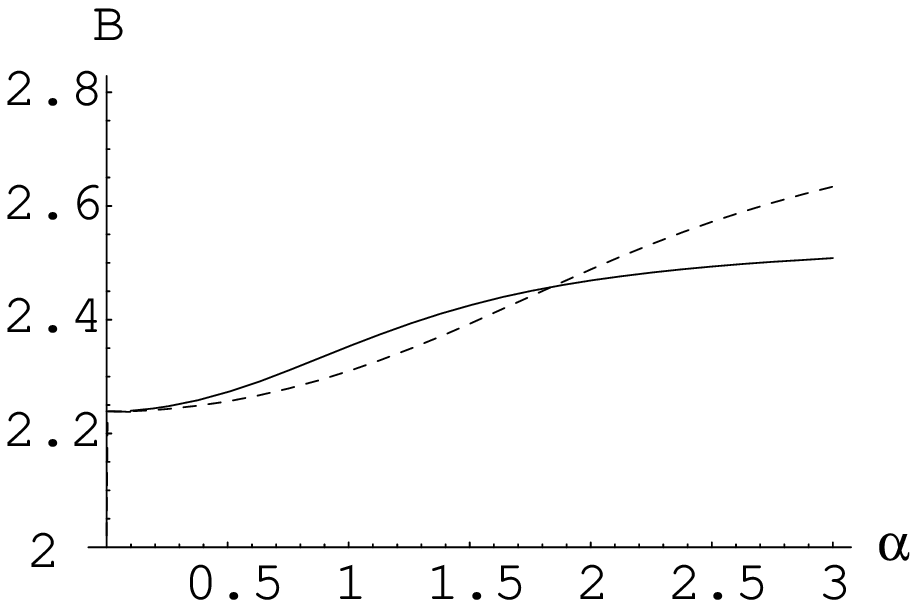}}}
\vspace{0.4cm}
\centerline{(b)\scalebox{0.75}{\includegraphics{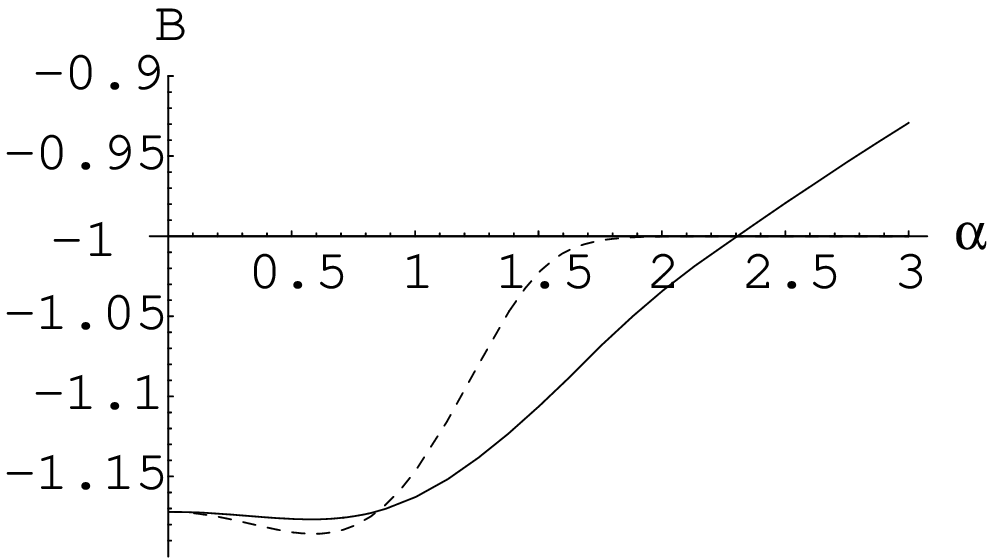}}}
\vspace{0.3cm}
\caption{
(a) The optimized Bell-CHSH function, ${\rm B}=|B_{CHSH}|_{max}$, for 
a PSGS (solid curve) and a SCS (dashed curve),
each of them divided at a 50:50 beam splitter,
with photon number parity measurements and the displacement operations.
(b) The optimized Bell-CH function ${\rm B}=|B_{CH}|_{max}$ for 
a PSGS (solid curve) and a SCS (dashed curve),
 each of them divided at a 50:50 beam splitter,
with photon   measurements and the displacement operations.
}
\label{fig:wq}
\end{figure}

The optimized Bell-CHSH functions, $|B_{CHSH}|_{max}$,
for the cases of the PSGS and the SCS are plotted in Fig.~\ref{fig:wq}(a).
In this figure, the SCS of amplitude $\alpha$ and the PSGS that best
approximates the SCS for the given amplitude are compared.
The PSGS is found to show larger
violations for $\alpha<1.84$, i.e., for the range where 
the fidelity between the PSGS and the SCS is $F>0.91$.
Note that the fidelity 
between the two states is high 
when $\alpha$ is small enough,
and in this limit (i.e. $\alpha\rightarrow0$) 
the optimized Bell-CHSH functions for the two states become
the same as show in Fig.~\ref{fig:wq}(a).

\subsection{Bell-CH inequality formalism with the $Q$ functions}

BW used  the $Q$ function for the test of the Bell-CH inequality violation 
using photon presence (i.e. on/off) measurements
\cite{BW}.
Note that in this paper, we refer to
a dichotomic measurement
discriminating between ``no photon" and ``any photon(s)"
as a photon presence measurement.
This is obviously more realistic for
an experimental Bell inequality test since
it is difficult to measure the parity of photon numbers
using currently available photodetectors.
The $Q$ function for a two-mode state $\rho_{12}$ is defined as
\begin{equation}
Q_{12}(z_1,z_2)=\frac{{}_2\langle z_2|{}_1\langle z_1|\rho_{12}
|z_1\rangle_1|z_2\rangle_2}{\pi^2},
\end{equation}
 where $|z_1\rangle$
and $|z_2\rangle$ are coherent states of amplitudes
$z_1$ and $z_2$ respectively.
The Bell-CH function in terms of $Q$ representation is 
\begin{equation}
\begin{aligned}
& B_{CH}
=\pi^2\big[Q_{12}(z_1,z_2)+Q_{12}(z_1,z_2^\prime)+Q_{12}(z_1^\prime,z_2) \\
&~~~~~~~~~~~~~~~~~~-Q_{12}(z_1^\prime,z_2^\prime)\big]
-\pi\big[Q_1(z_1)+Q_2(z_2)\big],
\label{ch-q}
\end{aligned}
\end{equation}
where
 $Q_1(z_1)$ and $Q_2(z_2)$  are the marginal $Q$ functions of modes
1 and 2. 
Then the local theory imposes the Bell-CH inequality
\begin{equation}
-1\leq
  B_{CH}\leq0.
  \end{equation}
Eq.~(\ref{ch-q}) is a generalized version of the BW's formalism
\cite{jeongsonkim}.
The $Q$ function for the entangled coherent state is
\begin{eqnarray}
&&Q_{ECS}(z_1,z_2)=N_-(\alpha)^2\Big\{
\exp[-|z_1-\beta|^2-|z_2+\beta|^2]\nonumber\\
&&~~~~~~~~~~~~~~~~~~~~~~~~~~+\exp[-|z_1+\beta|^2-|z_2-\beta|^2]\nonumber\\
&&~~-\exp[-(z_1-\beta)(z_1^*+\beta)
-(z_2+\beta)(z_2^*-\beta)-4\beta^2]\nonumber\\
&&~~-\exp[-(z_1^*-\beta)(z_1+\beta)
-(z_2^*+\beta)(z_2-\beta)-4\beta^2]
\Big\}.\nonumber\\
\label{qecs}
\end{eqnarray}
The marginal $Q$ function can also be calculated from (\ref{qecs}) so that
the Bell-CH function in Eq.~(\ref{ch-q}) can be obtained.
The optimized Bell-CH functions,
$|B_{CH}|_{max}$, for the cases of
the SCS of amplitude $\alpha$ and the PSGS optimized to approximate
the SCS are plotted in Fig.~\ref{fig:wq}(b).
Interestingly, the SCS shows large Bell violations when $\alpha<0.85$ ($F>0.999$)
but the PSGS outperforms the SCS when $\alpha>0.85$. 
This result is obviously different from the one for
the Bell-CHSH inequality using photon parity measurements
shown in Fig.~\ref{fig:wq}(a).

\section{
Analysis with impure Gaussian states, 
a beam splitter and a realistic avalanche photodetector}

In this section, we consider some detrimental factors that can
affect the Bell inequality tests with the PSGSs in real experiments.
First, experimental Gaussian states are not pure states
but have little amount of mixing. Therefore, we take
impurity of the squeezed states into consideration in our calculation.
Second, when one actually subtracts a photon from a Gaussian state,
a beam splitter with a high transmittivity and an avalanche photodetector,
which cannot resolve photon numbers, are typically used.
In such an experimental setup, if the transmittivity of the
beam splitter approaches 1, 
the probability of subtracting only one photon  becomes close to 100\%,
provided the detector clicked.
However, in this limit the success
probability (i.e., the probability of the ``click" event)
approaches zero. Therefore we need to consider 
the cases where the transmittivity of the beam splitter is not unity.
In this case, the generated state will be a mixed state
and that may affect the results of the Bell inequality tests.
Furthermore, we need to assess the effects of the non-unit efficiency 
and dark counts of the realistic avalanche photodetector used for photon subtraction.
Note that we shall consider the PSGSs of $r>0$ 
in this section without loosing generality.

\subsection{Effects of impure Gaussian states, 
a beam splitter and an ideal avalanche photodetector}

Kim {\it et al.} derived the characteristic function and the
Wigner function of the PSGS under experimentally realistic assumptions
as follows \cite{Kim05}.
We are interested in Gaussian states with the characteristic functions in the form of
\begin{equation}
C(\xi)=\exp\left(-{\frac{A}{2}}\xi_r^2-{\frac{B}{2}}\xi_i^2 \right),
\label{1}
\end{equation}
where $A$ and $B$ are determined by the quadrature variances of the field.
When the squeezed Gaussian field in Eq.~(\ref{1}) is 
divided at a beam splitter,
the two-mode characteristic function for the output field of
modes 1 and 2 is \cite{KimLee}
\begin{equation}
C_{out}(\eta,\xi)=\exp\left(-{1\over 2}{\bf x V}{\bf x}^T\right)
\label{character-out}
\end{equation}
where ${\bf x} = (\eta_r, \eta_i, \xi_r, \xi_i)$ and the
correlation matrix
\begin{equation}
{\bf V}=
\begin{pmatrix}
n_1 & 0 & c_1 & 0 \\
0 & n_2 & 0 & c_2 \\
c_1 & 0 & m_1 & 0 \\
0 & c_2 & 0 & m_2
\end{pmatrix}
\label{correlation}
\end{equation}
with 
\begin{eqnarray}
\begin{aligned}
&n_1=TA+R,~~n_2=TB+R,~~c_1=\sqrt{TR}(A-1),\\
&c_2=\sqrt{TR}(B-1),~~ m_1=RA+T,~~m_2=RB+T.
\end{aligned}
\label{value}
\end{eqnarray}

When the avalanche photodetector, which does not resolve photon numbers,
clicks at mode $2$, the generated state at mode $1$ is
\begin{equation}
\hat{\rho}_a={\cal N}\sum_{n=1}^{\infty}~_2\langle
n|\hat{\rho}_{out} |n\rangle_2 \label{any-rho},
\end{equation}
where ${\cal N}$ is the normalization factor.
Consider the unnormalized density operator for mode 1 of the output field
\begin{equation}
\hat{\rho}_t=\mbox{Tr}_2[\hat{\rho}_{out}]=\sum_{n=0}^{\infty}
~_2\langle n|\hat{\rho}_{out}|n\rangle_2. \label{rho-t}
\end{equation}
It is then clear from Eqs.(\ref{any-rho}) and (\ref{rho-t}) that
\begin{equation}
\hat{\rho}_a={\cal N}(\hat{\rho}_t-~_2\langle 0|\hat{\rho}_{out}
|0\rangle_2) \label{new-rho}
\end{equation}
where
\begin{equation}
\hat{\rho}_t={1 \over \pi}\int C_{out}(\eta,0)
\hat{D}_1(-\eta)d^2\eta
\end{equation}
and
\begin{equation}
_2\langle 0|\hat{\rho}_{out}|0\rangle_2= {1 \over \pi^2}\int
C_{out}(\eta, \xi)\mbox{e}^{-|\xi|^2/2}\hat{D}_1(-\eta)d^2\eta
d^2\xi. \label{102}
\end{equation}
Using $C_{out}(\eta,\xi)$ in Eqs.~(\ref{character-out}) to (\ref{value})
and Eqs.~(\ref{new-rho}) to (\ref{102}),
the characteristic function
 $C_a(\zeta)$ for $\hat{\rho}_a$ is found \cite{Kim05}:
\begin{eqnarray}
C_a(\zeta)&=& {\rm Tr}[\hat \rho_a \hat{D}(\zeta)] \nonumber \\
&=&{\cal N}\mbox{e}^{-{1\over
2}(n_1\zeta_r^2+n_2\zeta_i^2)} \Big[1 -
\frac{2}{\sqrt{(m_1+1)(m_2+1)}}\nonumber \\
&~&\times\exp\Big(\frac{c_1^2}{2(m_1+1)}\zeta_r^2
+ \frac{c_2^2}{2(m_2+1)}\zeta_i^2\Big)\Big]. \label{103}
\end{eqnarray}
The normalization factor is calculated as
$$
{\cal N}=\frac{\sqrt{(m_1+1)(m_2+1)}}
{\sqrt{(m_1+1)(m_2+1)}-2}.
$$
It is then straightforward to derive the Wigner function 
from Eq.~(\ref{103}) by the Fourier transform as
performed in Eq.~(\ref{koala}). 
The $Q$ function is obtained in the same way 
but using the normally ordered
characteristic function $C^Q_a(\zeta)=C_a(\zeta)\exp[-(1/2)|\zeta|^2]$.
We also obtain
the success probability of the ``click" event
for the avalanche photodetector as
\begin{equation}
P_s=1-{}_2\langle 0| 
\rho_t^\prime
|0\rangle_2
=\frac{2}{\sqrt{(1+m_1)(1+m_2)}}
\end{equation}
where
\begin{equation}
\hat{\rho}_t^\prime={1 \over \pi}\int C_{out}(0,\xi)
\hat{D}_2(-\xi)d^2\eta. 
\end{equation}

\begin{figure}
\centerline{\scalebox{0.7}
{~~(a)~~~~\hspace{0.4cm}\includegraphics{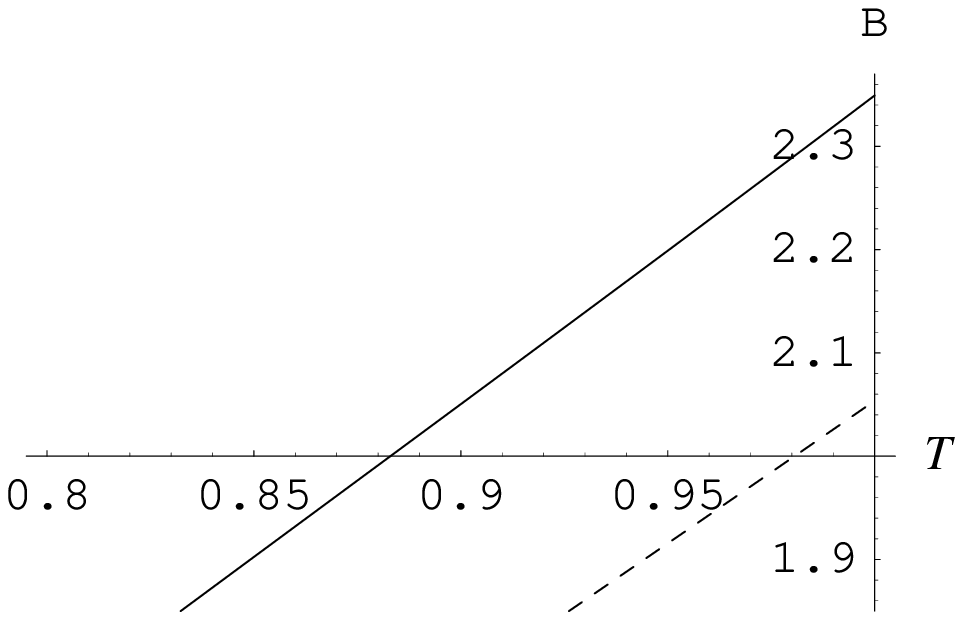}}~~~~~}
\vspace{0.4cm}
\centerline{\scalebox{0.8}{(b)\includegraphics{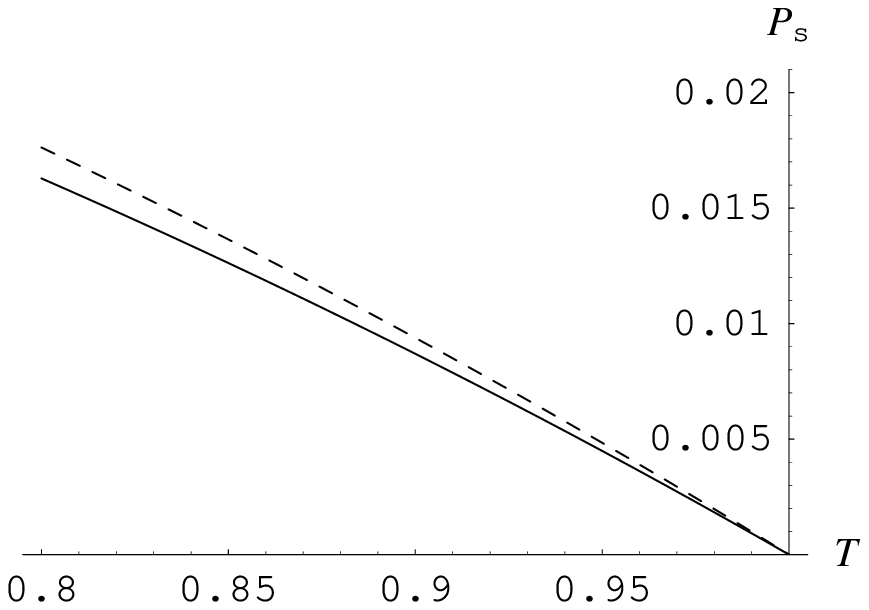}}}
\vspace{0.3cm}
\caption{
(a) The numerically optimized Bell-CHSH function, ${\rm B}=|B_{CHSH}|_{max}$,
for a PSGS using a pure Gaussian state
of $r=0.3$ ($2.61$dB, solid curve) 
and mixed Gaussian states (dashed curve).
The mixed Gaussian state corresponds to
$-2.56$dB below and $2.65$dB above
the vacuum level (the two dimensional variance $A\times B$ is about 1.02).
The horizontal axis represents 
the transmmittivity, $T$,
of the beam splitter used to generate
the PSGS.
(b) The success probability of
 the pure (solid curve) and mixed (dashed curve) cases
 shown in (a). 
}
\label{fig:w}
\end{figure}

The numerically optimized Bell-CHSH functions
are plotted and compared with the corresponding
success probabilities
in Fig.~\ref{fig:w}(a).
The violation for the pure PSGS ($r=0.3$)
decreases as $T$ gets smaller.
We also consider the total variance, $A\times B$,
which is strictly related to the purity in the case of Gaussian states,
as a measure of mixedness for the input Gaussian state.
The total variance is 1 for a pure  Gaussian state
and it increases as the level of mixedness becomes larger.
We can observe that the violation is quite sensitive
to the degree of mixedness. 
We note that only a small amount of mixing (total variance
about 1.02) decreases the
Bell violation significantly as shown in Fig.~\ref{fig:w}(a).
The success probablities for the corresponding cases are
plotted in Fig.~\ref{fig:w}(b).
When we equally increase $A$ and $B$,
the violation disappears when the variance around $A=1.024$.

\begin{figure}
\centerline{\scalebox{0.7}{(a)\includegraphics{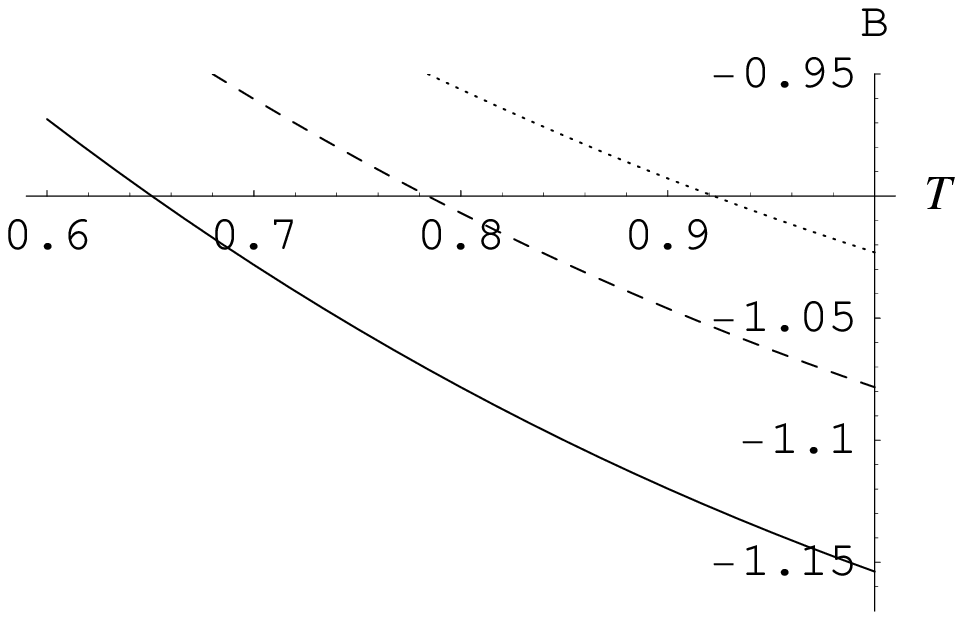}}}
\vspace{0.35cm}
\centerline{\scalebox{0.7}{(b)\includegraphics{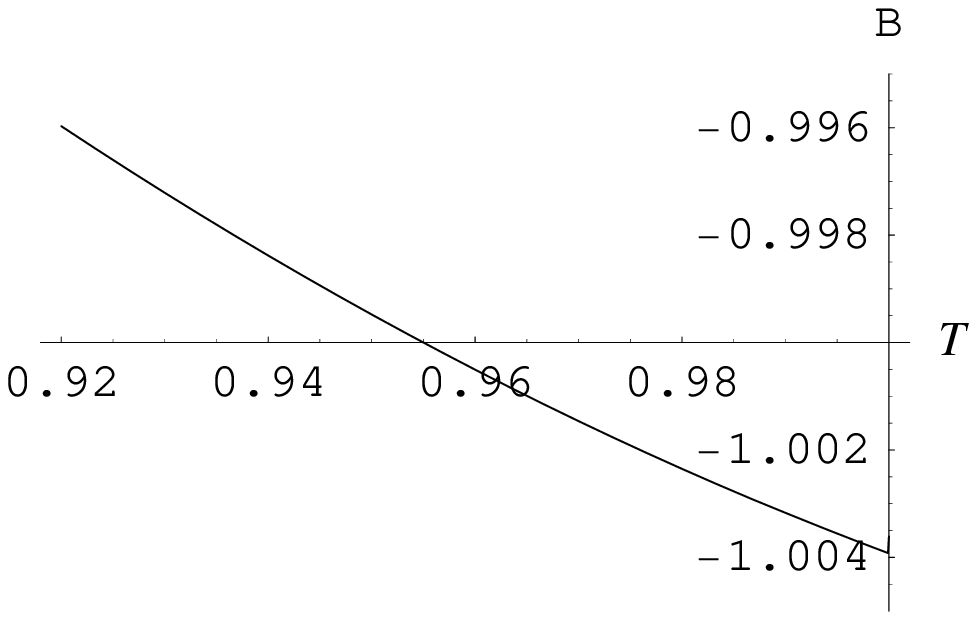}}}
\vspace{0.35cm}
\centerline{\scalebox{0.76}{~~~(c)\includegraphics{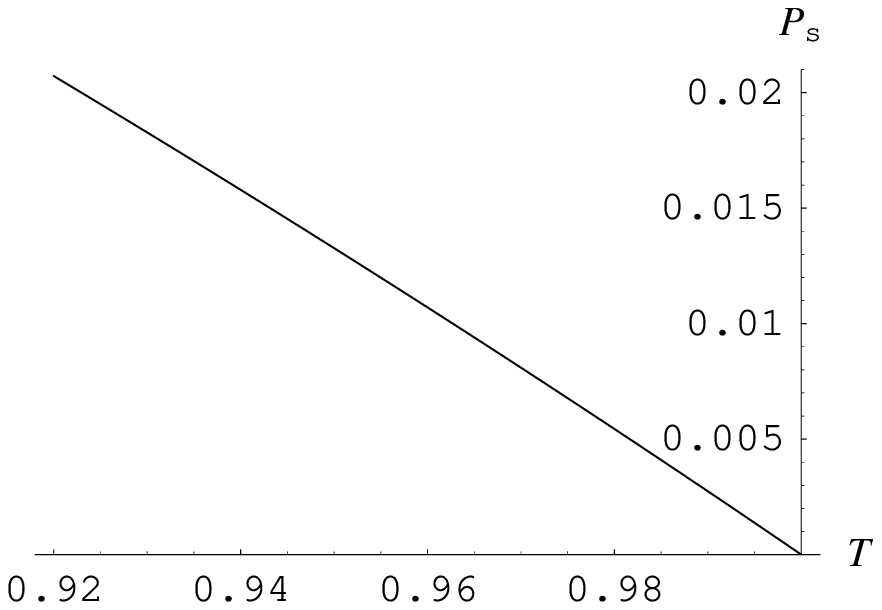}}}
\vspace{0.3cm}
\caption{
(a) The numerically optimized Bell-CH function, ${\rm B}=|B_{CH}|_{max}$,
for a PSGS using a pure Gaussian state
of $r=0.3$ ($2.61$dB, solid curve)
and mixed Gaussian states.
The mixed state case represented by
the dashed curve corresponds to
$-2.52$dB below and $2.69$dB above
the vacuum level (the variance $A\times B$ is about 1.04).
The case represented by
the dotted curve corresponds to
$-2.43$dB below and $2.78$dB above
the vacuum level (the variance $A\times B$ is about 1.08).
(b)
The numerically optimized Bell-CH function
for a PSGS
using an experimentally feasible Gaussian state.
The squeezing degree is $-3.57$dB below
and $4.26$dB above the vacuum level.
The two dimensional variance, $A\times B$, in this case is about $1.17$.
(c) The success probability $P_s$ of the case
shown in (b).
}
\label{fig:q}
\end{figure}

The numerically optimized Bell-CH functions
using photon presence measurements
are plotted in Fig.~\ref{fig:q}.
In this case, the violation is obviously less 
sensitive to the mixedness of the input Gaussian state.
We have considered a recent experimental achievement, where
the squeezing degree is $-3.57$dB below
and $4.26$dB above the vacuum level, and
the Bell-CH inequality is still violated \cite{kt4,Kentaro}.
The total variance in this case is about $1.17$.
This confirms that the Bell-CH inequality test
using photon presence measurements is experimentally
more feasible than the Bell-CHSH inequality
using photon number parity measurements.

Here we briefly compare our results with previous ones~\cite{In05}.
In our numerical study, the CH inequality using on/off photon detection 
is violated for $T>0.64$
when $r=0.39$  and $\eta=1$.
This is consistent with the results in Ref.~\cite{In05},
where the CHSH inequality based on the photon on/off measurements is analyzed.
However, in our study, the Bell-CHSH inequality using photon parity measurements
shows Bell-CHSH violations for $T>0.89$, while 
$T>0.8$ is sufficient for the Bell-CHSH violation based
on the photon on/off measurements.

\subsection{Effects of non-unit quantum efficiency
and dark counts for a realistic avalanche photodetector}

We need to consider the effects of non-unit quantum efficiency
of the avalanche photodetector used for photon subtraction.
The postselection to generate PSGS is made
only when the detector ``clicks" regardless of the number of photons.
As we pointed out, when the transmmittivity $T$ of the beam splitter is large,
the probability of subtracting only one photon
becomes dominant compared to the probability of subtracting more than one. 
This essential nature does not change even when the detection efficiency is low.  
Therefore, one may predict that the
Bell inequality tests with the PSGS will be insensitive to 
the detection inefficiency when $T$ is sufficiently large.
This is confirmed by the following analysis.

\begin{figure}
\centerline{\scalebox{0.8}{(a)\includegraphics{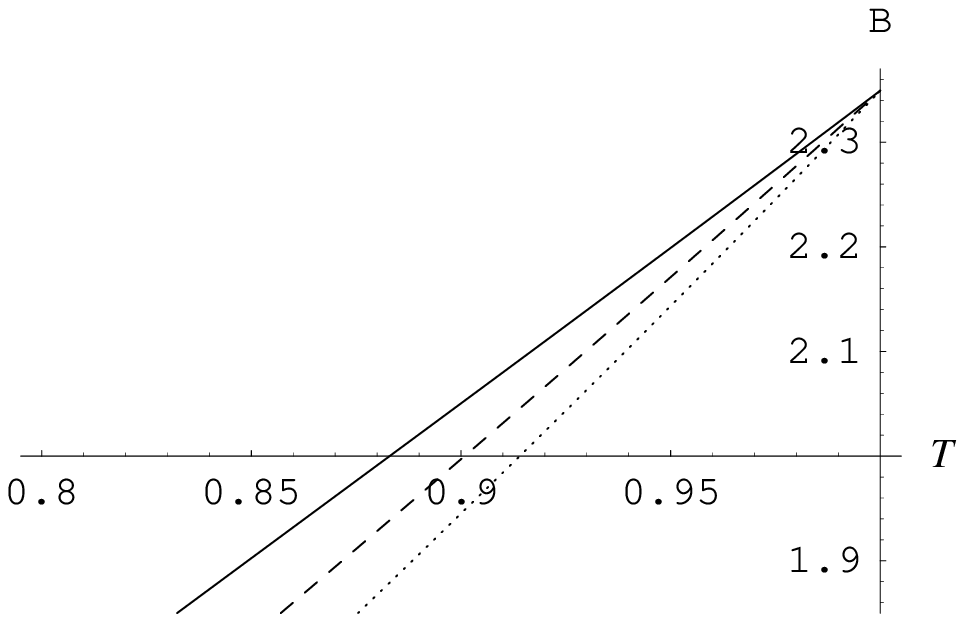}}}
\vspace{0.35cm}
\centerline{\scalebox{0.85}{(b)\includegraphics{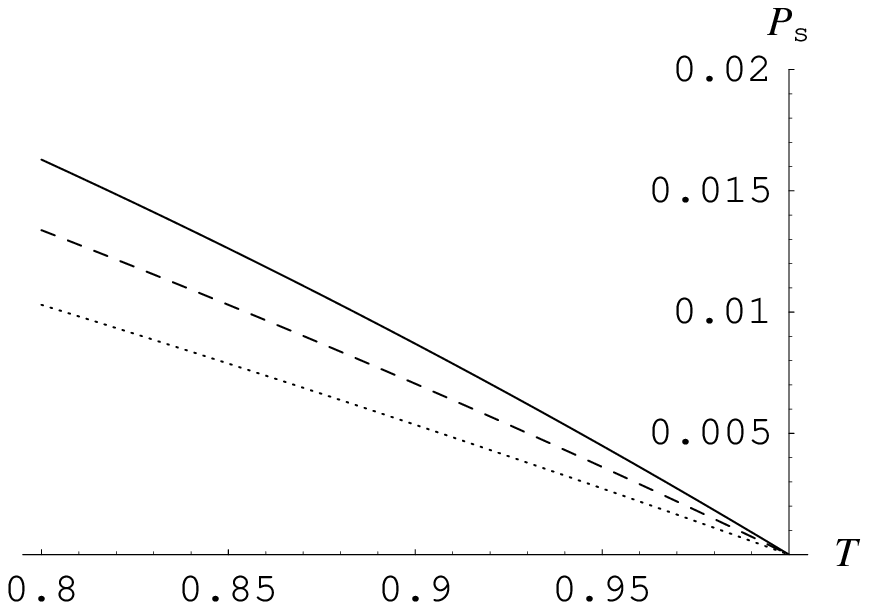}}}
\vspace{0.3cm}
\caption{(a) The numerically optimized Bell-CHSH function,
${\rm B}=|B_{CHSH}|_{max}$, for a PSGS using a pure Gaussian state
of $r=0.3$ with a beam splitter of transmmittivity $T$ and a
realistic detector of efficiency $\epsilon$.
The quantum efficiency of the detector is considered to be
$\epsilon=1$ (solid line), $\epsilon=0.8$ (dashed line)
and $\epsilon=0.6$ (dotted line).
(b) The success probability $P_s$ for the cases
shown in (a).}
\label{fig:rev}
\end{figure}

\begin{figure}
\centerline{\scalebox{0.8}{(a)\includegraphics{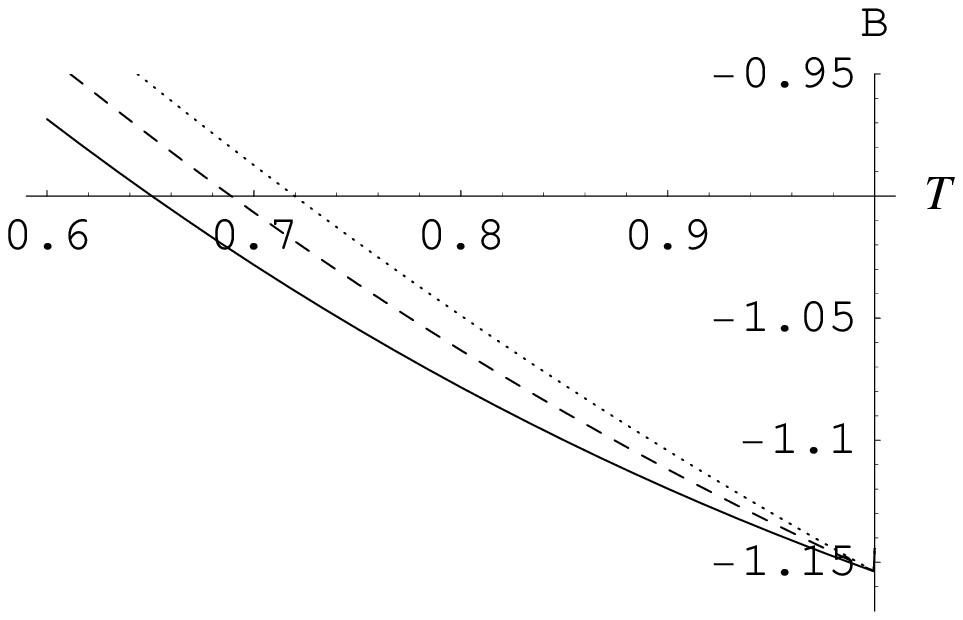}}}
\vspace{0.3cm}
\centerline{\scalebox{0.8}{(b)\includegraphics{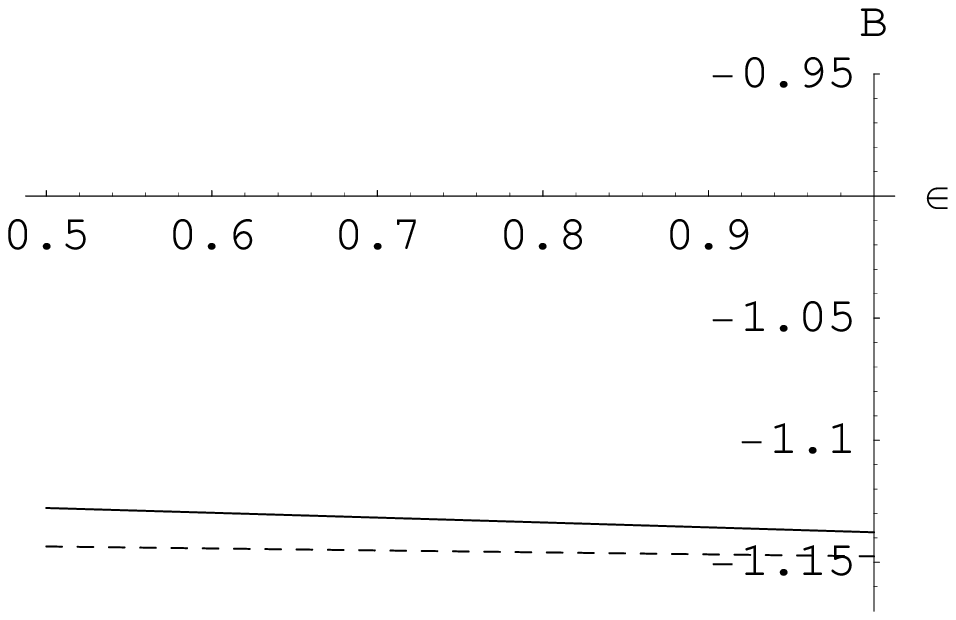}}}
\vspace{0.3cm}
\caption{(a) The numerically optimized Bell-CH function,
${\rm B}=|B_{CH}|_{max}$, for a PSGS using a pure Gaussian state
of $r=0.3$ with a beam splitter of transmmittivity $T$ and a
realistic detector of efficiency $\epsilon$.
The quantum efficiency of the detector is considered to be
$\epsilon=1$ (solid line), $\epsilon=0.8$ (dashed line)
and $\epsilon=0.6$ (dotted line).
(b) The numerically optimized Bell-CH function,
${\rm B}$, against the detection efficiency $\epsilon$
for $T=0.95$ (solid line) and $T=0.98$ (dashed line).
}
\label{fig:Qrev}
\end{figure}

Olivares and Paris obtained the generalized Quasi-probability function $W_{j}(z)$
of the photon subtracted squeezed state with
the detection efficiency $\epsilon$
as \cite{OP}
\begin{equation}
W_{j}(z)=N_\epsilon\Big({\cal G}_1(z)-\frac{{\cal G}_2(z)}
{\epsilon\sqrt{{\rm Det}[H+\sigma_{\rm M}]}}\Big)
\end{equation}
where $N_\epsilon$ is the normalization factor and
\begin{equation}
\begin{aligned}
&{\cal G}_k(z)=\frac{2\exp[-\frac{2(2{\cal A}_k-j)|z|^2+4{\cal B}_k(z^2+{z^*}^2))}
{(2{\cal A}_k-j)^2-16{\cal B}_k^2}]}{\pi\sqrt{(2{\cal A}_k-j)^2-16{\cal B}_k^2}},\\
&H=\left(\begin{array}{cc}
h^+ & 0  \\
  0 & h^-
\end{array}\right), \\
&\sigma_M=\frac{2-\epsilon}{2\epsilon}\openone,
\end{aligned}
\end{equation}
with
\begin{equation}
\begin{aligned}
&{\cal A}_k=\frac{1}{2}(a^+_k+a^-_k),\\
&{\cal B}_k=\frac{1}{4}(a^-_k-a^+_k),\\
&a^\pm_1=\frac{1}{2}\{1+(e^{\pm 2r}-1)t\},\\
&a^\pm_2=\frac{1}{2}\pm \frac{t \sinh r }{\cosh r\mp \{1-\epsilon(1-t)\}\sinh r},\\
&h^\pm=\frac{1}{2}\{e^{\pm 2r}(1-t)+t\}.
\end{aligned}
\end{equation}
Here, the function $W_j(z)$ becomes the Wigner function when $j=0$,
while it becomes $Q$ function when $j=-1$.
It is then straightforward to obtain 
the Bell-CHSH function and the Bell-CH function
using the Wigner function and $Q$ function,
respectively, using Eqs.~(\ref{eq:BF}) and (\ref{ch-q}).    
We have plotted the numerically optimized
Bell-CHSH and the Bell-CH functions
in Figs.~\ref{fig:rev} and \ref{fig:Qrev}, respectively.
It shows again that 
the Bell-CH inequality is more robust against detection inefficiency. 
In Fig.~\ref{fig:Qrev}(b),
the numerically optimized Bell-CH function
is plotted against the detection efficiency $\epsilon$
for $T=0.95$ (solid line) and $T=0.98$ (dashed line).
It seems obvious that when the transmmittivity $T$ is as large as $T\geq0.95$,
detrimental effects of the detection inefficiency are very small.

\begin{figure}
\centerline{\scalebox{0.8}{\includegraphics{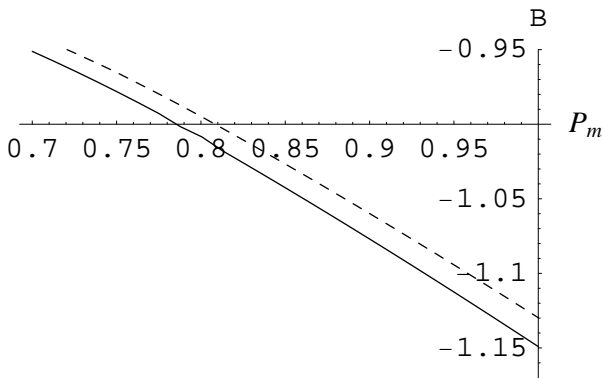}}}
\vspace{0.3cm}
\caption{The numerically optimized Bell-CH function,
${\rm B}=|B_{CH}|_{max}$, for a PSGS using a pure Gaussian state
of $r=0.3$ 
against the modal purity factor $P_m$ with
a beam splitter of transmmittivity $T=0.99$ (solid line)
and $T=0.95$ (dashed line). The detection efficiency was supposed to
be $\epsilon=0.6$.
}
\label{fig:dark}
\end{figure}

If the dark count rate of the photodetector used to subtract a photon 
is non-negligible, the resulting state will be in a mixture of the 
photon subtracted squeezed state and the squeezed vacuum.
Such a mixed state can be represented as \cite{Kim05,kt1}
\begin{equation}
P_m W(z)+(1-P_m)W_{s}(z)
\end{equation}
where $W(z)$ is the Wigner function of the photon subtracted squeezed state,
$W_{s}(\alpha)$ is the Wigner function of the squeezed vacuum and
$P_m$ is called the modal purity factor.
The numerically optimized Bell-CH function
against the modal purity factor is plotted in Fig.~\ref{fig:dark}.
If $P_m\lesssim 0.78$, the Bell violation cannot be observed
even though $T$ is as large as $T=0.99$
and the initial squeezed state was pure.

\section{remarks}

We have studied tests of Bell inequalities with PSGSs.
It has been found that the PSGSs  
largely violate the Bell-CHSH inequality using
photon number parity measurements and
the Bell-CH inequality using photon presence measurements.
The PSGSs and the SCSs
violate Bell inequalities 
in different manners.

We have analyzed the effects of the key experimental components used to
generate the PSGSs:
the mixedness of the Gaussian states,
limited transmittivity of the beam splitter 
and the avalanche photodetector which cannot
resolve photon numbers have been taken into consideration
 in our analysis.
As a result of this analysis, the degrees
of mixedness and the beam
splitter transmittivity that can be allowed for successful tests of Bell's
inequality have been revealed. 
The tests of the Bell-CH inequality using photon presence measurements
have been found to be less sensitive to the  
mixedness of the Gaussian state used to generate the PSGS.
We have also analyzed the effect of the non-unit quantum efficiency and
dark counts of the photon detector used for photon subtraction.
We have pointed out that when the transmmittivity $T$ is large ({\it e.g.} $T\geq0.95$),
detrimental effects of the detection inefficiency are very small.

We now address experimental feasibility of the Bell inequality tests discussed
in this paper.  
The photon presence measurements are obviously easier to realize
using current technology, and therefore the Bell-CH inequality is
a better candidate for nonlocality tests of the PSGSs.
Of course, the Bell-CH inequality tests using photon presence measurements
may still suffer the inefficiency of the photon detectors used
for Bell inequality tests (not the one used for photon subtraction)
in real experiments. This is a nontrivial problem to overcome.

As a less challenging experimental task, one may consider the tomography approach
of verification of quantum nonlocality using homodyne detection 
used in Ref.~\cite{Z2006}.
Our study of the Bell inequality tests
is based on the quasi-probability functions \cite{BW}, and
it is possible to reconstruct the quasi-probability
functions using homodyne detection.
Homodyne detection can be highly efficient using current technology
\cite{kt4}.
Furthermore, even when the homodyne efficiency is not satisfactory,
one may always correct losses at the detectors to reconstruct
the quasi-probability functions of the {\it generated} state.
The reconstruction of the two-mode and single-mode $Q$ functions
would be the most efficient method to verify quantum nonlocality
 of the generated PSGS.
If the currently available Gaussian state is used and the final
homodyne inefficiency is corrected,
one can experimentally show that the generated state
is the one that violates the Bell-CH inequality,
as predicted in Fig.~\ref{fig:q}.
Even though such a method is not a direct test of Bell's inequality
using homodyne detection described in Refs.~\cite{Nha,G04,Magda},
which are experimentally more demanding,
it is enough to show that quantum mechanically
nonlocal states have been generated \cite{Z2006}.

Of course, it should be noted that the violation of the Bell-CH inequality 
shown in Fig.~\ref{fig:q}(b), based on a recent experimental achievement,
is small. Such small values may disappear
by other experimental factors such as the dark count rate of
the avalanche photodetector.
This indicates that it is important to improve the purity of the
Gaussian state to clearly verify quantum nonlocality of the two-mode PSGS
in real experiments.

\acknowledgments

This work was supported by the
 DTO-funded U.S. Army Research Office Contract No. W911NF-05-0397, the 
 Australian Research Council and the Queensland State Government.
The author thanks Kentaro Wakui for his useful comments on his
recent experiment \cite{kt4,Kentaro}.

\end{document}